# A rapid-onset diffusion functional MRI signal reflects neuromorphological coupling dynamics

Daniel Nunes, Rita Gil, Noam Shemesh[*]

*Champalimaud Research, Champalimaud Centre for the Unknown, Lisbon, Portugal*

**\*Corresponding author:**
Dr. Noam Shemesh, Champalimaud Neuroscience Programme, Champalimaud Centre for the Unknown, Av. Brasilia 1400-038, Lisbon, Portugal**.**
E-mail: noam.shemesh@neuro.fchampalimaud.org .
Phone number: +351 210 480 000 ext. #4467.

**Running title:** Diffusion weighted fMRI tracks neuromorphological coupling dynamics

**ORCID**: 0000-0001-6681-5876 (NS)

**Author Contributions**
NS designed research, contributed new analytic tools, analyzed data and wrote the paper. DN and RG performed research, contributed new analytic tools, analyzed data and designed research and contributed to writing the paper. All authors proofread the paper.

**Competing interests: None.**

**This PDF file includes:**
    Main Text
    Figures 1 to 4
    Tables N/A

# Abstract

Functional Magnetic Resonance Imaging (fMRI) has transformed our understanding of brain function in-vivo. However, the neurovascular coupling mechanisms underlying fMRI are somewhat "distant" from neural activity. Interestingly, evidence from Intrinsic Optical Signals (IOSs) indicates that neural activity is also coupled to (sub)cellular morphological modulations. Diffusion-weighted functional MRI (dfMRI) experiments have been previously proposed to probe such *neuromorphological couplings*, but the underlying mechanisms have remained highly contested. Here, we provide the first direct link between in vivo ultrafast dfMRI signals upon rat forepaw stimulation and IOSs in acute slices stimulated optogenetically. We reveal a hitherto unreported rapid onset (<100 ms) dfMRI signal which (i) agrees with fast-rising IOSs dynamics; (ii) evidences a punctate quantitative correspondence to the stimulation period; (iii) and is rather insensitive to a vascular challenge. Our findings suggest that neuromorphological coupling can be detected via dfMRI signals, auguring well for future mapping of neural activity more directly.

# Introduction

Functional Magnetic Resonance Imaging (fMRI)[1,2] has transformed neuroscience and biomedicine by enabling a noninvasive vista into global brain activity in health and disease. To deliver contrasts reflecting brain activation, fMRI typically harnesses neurovascular couplings linking neural activity with blood oxygenation, blood flow, and blood volume dynamics. Upon stimulus presentation, or even at resting state, these so-called "Blood Oxygenation Level Dependent" (BOLD) neurovascular coupling mechanisms[1–7] produce local magnetic susceptibility modulations which can be detected using MRI[8]. Such susceptibility-driven fMRI methods have been instrumental for understanding brain function upon task presentation[9], at rest[10], upon learning[11,12], in disease[13,14], and for generating computational models of brain activity[14,15].

Despite much effort[5,16–19], the exact nature of neurovascular couplings remains elusive, thereby complicating the interpretation of fMRI experiments[20]. In particular, hemodynamic responses are surrogate indicators of neural activity, both spatially and temporally. Local neural activity can recruit vasculature from distant areas thereby decreasing the spatial specificity of fMRI[21]. fMRI's temporal specificity is typically limited by the slowness ensuing hemodynamics[5], although recent measurements with high temporal resolution have revealed that BOLD rise times are quite representative of underlying neural activity when measuring cortical activity[22,23]. Together with the stringent requirements for physiological stability, the above mentioned limitations of the BOLD mechanism have prompted the development of other sources of functional contrasts for mapping brain activity using magnetic resonance, including: neurotransmitter-specific contrast agents[24]; methods measuring correlates of electrical properties[25]; and diffusion fMRI[26].

Diffusion MRI imparts sensitivity to micron-scale displacements on the measured signal[26]. Fortuitously, cellular scale morphologies and boundaries modulate the distance travelled by water molecules on a typical dMRI timescale[27] (several milliseconds, typically), making the diffusion-weighted MRI signal a powerful probe of microstructure[28] and structural connectivity[29]. In 2001, Darquie et al. observed stimulus-locked dynamics in diffusion-weighted fMRI (dfMRI) signals[30], and later studies have suggested that dfMRI therefore detects "cell swelling" mechanism coupled with neural activity and thereby report on activity more directly[31–34]. Indeed, pioneering diffusion fMRI (dfMRI) measurements[31–34] exhibited several notable features: (i) a more focal localization of "active" areas compared to BOLD counterparts; (ii) a resilience to BOLD-suppressing pharmacological agents; (iii) an arguably faster response compared to BOLD counterparts. Although some of these features have been reproduced in rodents[32,33,35] and acute tissue preparations[36,37], others have vigorously claimed that dfMRI contrasts were merely "filtered" BOLD responses. Miller et al. noticed strong dfMRI signal modulations upon hypercapnia, suggesting the contribution of a prominent vascular component, and pointed out that dfMRI signals were not as fast as previously thought[38]. The more focal dfMRI activation areas were attributed to the inherently lower dfMRI signal-to-noise ratio[39]. The controversy was further intensified by studies investigating different echo times and diffusion weightings[40,41], as well as multimodal experiments in brain slices at low field[42], suggesting that the contrasts observed in dfMRI are unlikely to be of a neural origin.

Temporally coupled radial displacements have been observed in isolated nerve preparations upon action potential passage, with millisecond accuracy[43], suggesting a strong coupling between neural activity and microscopic morphological features (hereafter referred to as the neuromorhological coupling). Intrinsic optical signals (IOSs) showed that action potential firing is associated with cellular volume changes due to water influx through the cell membrane[44] and shrinkage of the interstitial volume[45]. Additionally, IOSs were shown to have a significant neuronal and astroglial contributions[46].

Interestingly, these microstructural biological processes are typically characterized by a wide dynamic range, including a rapid (millisecond-scale) onset[43], and a slow (10s of seconds) return to baseline[47]. Here, we hypothesized that dfMRI may capture some of these temporal signatures, particularly the more rapid cell volume changes and extracellular volume alterations coupled to neuronal activity. To assess this, high-temporal resolution dfMRI experiments are required. To our knowledge, all dfMRI experiments have been performed with low temporal resolution, typically in the range of 1- 3 sec, thereby obscuring putative rapid microstructural dynamics with BOLD responses. To shed light into the nature of dfMRI signals, we developed an ultrafast line-scanning dfMRI approach, which enabled the investigation of dfMRI signals with a temporal resolution of 100 ms. Our findings in rat somatosensory cortex suggest that dfMRI signals contain (at least) two different components: a rapid-onset component, which was rather insensitive to hypercapnia, suggestive of a neuromorpholgical coupling, and another slower component more sensitive to hypercapnia, suggesting more neurovascular origins. Intrinsic optical microscopy (IOM) time-series in optogenetically-stimulated acute brain slices corroborated a close similarity between fast rise-time IOSs and the rapid-onset dfMRI component, in turn suggesting that dfMRI can reflect, at least in part, a neuromorphological coupling.

# Results

To measure in vivo dfMRI and spin-echo BOLD (SE-BOLD) fast dynamics, we performed forepaw stimulation (Figure 1A) and recorded the signals in the forelimb primary somatosensory cortex (FL S1) of rats, using a large-tip-angle[48] line-scanning[22] (LTA-LS) approach (Figure 1B-D).

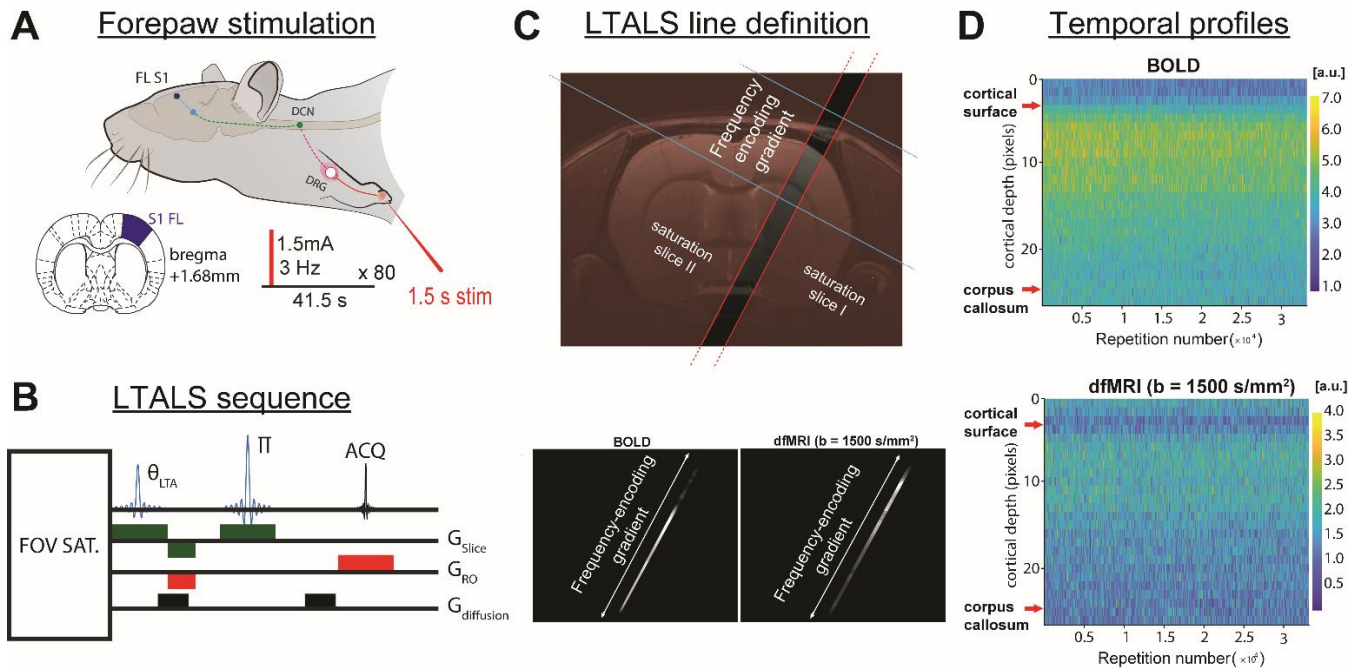

**Figure 1.** Large tip-angle line scanning data acquisition. **(A)** Schematic representation of the stimulation paradigm in sedated rats. Stimulation needles in the left forepaw delivered electrical square pulses to induce temporally precise activation of the forelimb primary somatosensory cortex (FL S1). **(B)** Schematic representation of the pulse sequence used to acquire LTA-LS data. A pulsed gradient spin-echo sequence compromised a saturation band module, a tip-angle $\theta_{LTA}$=155 degrees, followed by a $\pi$ pulse. A slice selective gradient ($G_{slice}$) was used to spatial encoding and readout gradients ($G_{RO}$) for frequency encoding during acquisition (ACQ). Diffusion sensitizing gradients ($G_{diffusion}$) were applied on each side of the $\pi$ pulse to impart diffusion weighting in dfMRI measurements and were turned-off during SE-BOLD measurements. **(C)** Diagram of the line scanning method, showing the positioning of the saturation slices.
Abbreviations: DRG – dorsal root ganglion; DCN – dorsal central nucleus (dorsal column-medial lemniscus pathway); FL S1 – forelimb primary somatosensory cortex.

Maps corresponding to line profiles scanned every 100 ms in the rat somatosensory cortex are presented in Figure 2. The top row (Figure 2A) represents SE-BOLD dynamics; the middle row (Figure 2B) depicts diffusion weighted fMRI dynamics at b = 1.5 ms/µm²; and the bottom row (Figure 2C) represents the temporal profile of the quantitative apparent diffusion coefficient calculated from the other two signals (c.f. Eq. 1). Figures 2D-F reflect the spatially averaged corresponding signals. The SE-BOLD dynamics (Fig. 2A) onset around 1.3 sec, and peak between 2 to 3 seconds, depending on the cortical layer.

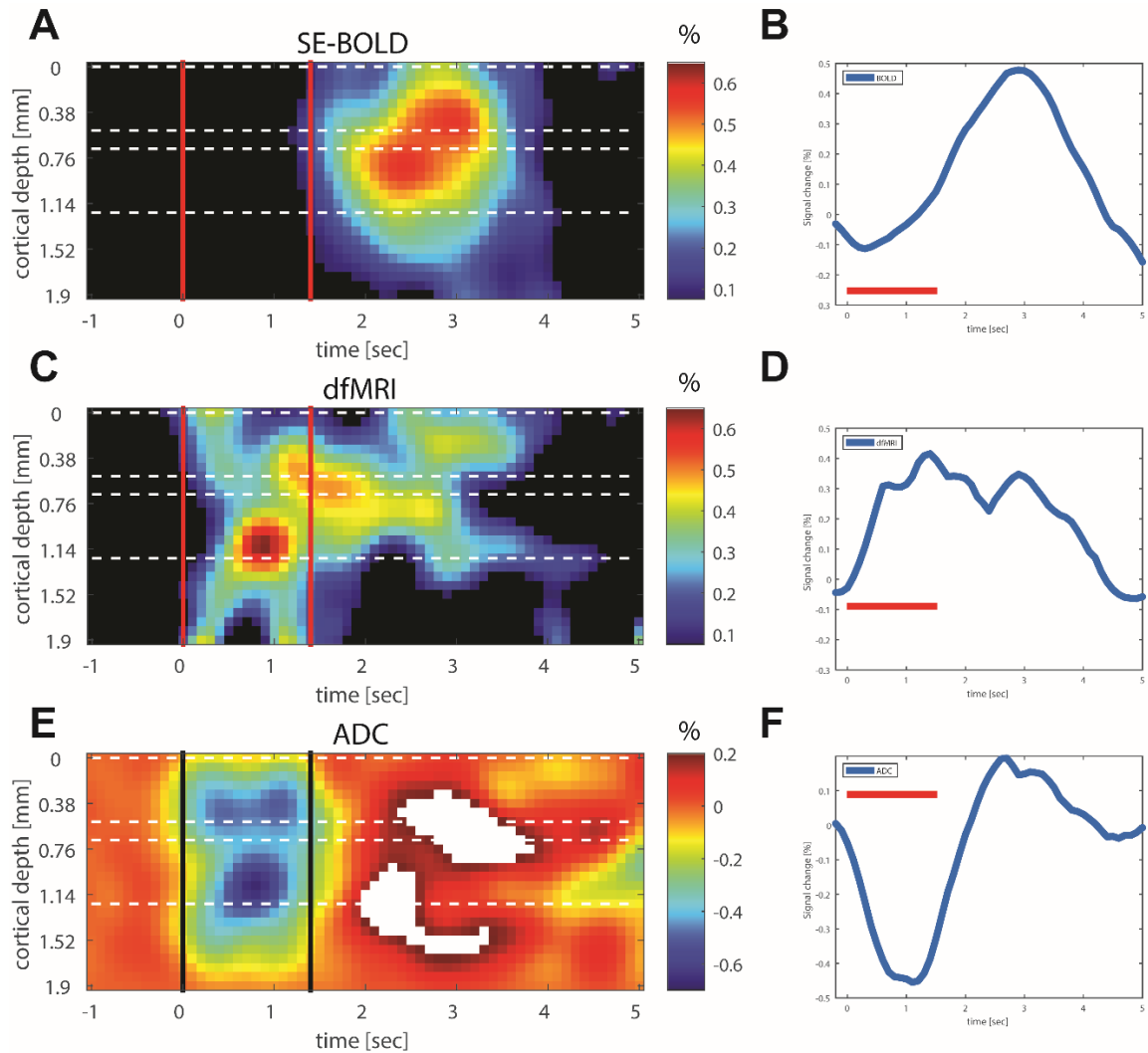

**Figure 2.** Temporal profiles of FL S1 activation. **(A)** SE-BOLD, **(B)** dfMRI and **(C)** ADC temporal activation maps, in percentage, of the FL S1 region. In the left, the cortical depth denotes the FL S1 column depth and in the right cortical layers I to VI are denoted, highlighted by white dashed horizontal lines. The red vertical solid bars denote stimulation time. **(D)** SE-BOLD, **(E)** dfMRI and **(F)** ADC spatially averaged time courses of the data, in percentage change. The red horizontal bars denote stimulation. Note the fast, nearly immediate rise of dfMRI signals upon stimulation, and the temporally punctate ADC response to the stimulation.

Remarkably, the dfMRI dynamics are completely different: already at the first measurement following the stimulus onset (100 ms), dfMRI signals exhibit higher signal than in the rest period. Layer V dfMRI signals appear slightly later, around ~300 ms after the beginning of the forepaw stimulation. A first peak is observed in layers II/III and IV about 300-500 ms after stimulation onset, followed by a large peak in layer V at 800-1000 ms. These signals then start decreasing in amplitude, followed by new peaks and dynamics around 1300 ms for layer II/III and between 2000-3000 ms for deeper cortical layers. These later dfMRI signals exhibit some latency until around 4.5 seconds post stimulation. In other words, *two* components can be generally observed in dfMRI signals: one with a rapid onset, and another with slower dynamics. Interestingly, these early dfMRI responses exhibited relatively invariant properties when experiments were repeated under a hypercapnia challenge (Figure S1); the later dfMRI components modulate more strongly upon a hypercapnic challenge. BOLD responses were modulated between hypercapnic and normal conditions in the first 3 seconds following stimulation.

To further dissect these signals, Figure 2C plots the behavior of the quantitative apparent diffusion coefficient (ADC) over time. The ADC dynamics are different than those of each of the components shown in Figs. 2A and 2B. Although the ADC signals are also characterized by a rapid onset (mirroring that of dfMRI-weighted signals), a perhaps more striking feature of the ADC signals is that they appear highly temporally punctate; note that the signal modulations associated with negative changes in ADC occur nearly only during the stimulation epoch, and, following the rapid onset, the negative ADC signal change ceases nearly as soon as the stimulation ends.

We decided to further investigate the nature of these fast dfMRI signals and their relationship to microstructural dynamics by performing auxiliary IOM experiments where (1) the signals are well-known to correspond to microstructural effects and (2) BOLD effects are absent. To enable temporally precise control over neural activity, these IOM experiments were performed using acute brain slices harvested from knock-in mice expressing the light activated ion channel channelrhodopsin-2, fused with yellow fluorescent protein (ChR2-YFP) under the control of the Thy1-promoter, making them amenable to optogenetic manipulation. A setup compatible with simultaneous optogenetics and intrinsic optical microscopy was built (Figure 3A) and the functional experiment was performed with a single stimulation epoch (1 sec stimulation duration at 20Hz, with a pulse width of 10msec and a laser power of 5 mW, Figure 3B). Figures 3C-D present fluorescence microscopy in these slices, confirming the expression of ChR2-YFP in the hippocampus of Thy1$^{+/-}$ transgenic mice; bright field microscopy (Figures 3E-F) is shown for anatomical comparison. Figure 3G shows a representative image from the functional IOM experiment time series, revealing high resolution and signal to noise in the slice. When ROIs were drawn over the entire hippocampus, IOSs were clearly observed for slices extracted from Thy1$^{+/-}$ transgenic mice (red trace). Identical experiments performed on control slices (C57Bl6 wild-type mice), showed no sign of activation (blue trace), as expected, but confirmed the stability of the IOM setup (Fig. 3G). Finally, to confirm that the observations are not specific to hippocampus, another set of experiments was performed on acute cortical slices from the same mouse line, expressing ChR2 mainly in layer V neurons in the cortex. The results clearly show that the rise times are very similar for both specimens (Fig. 3G).

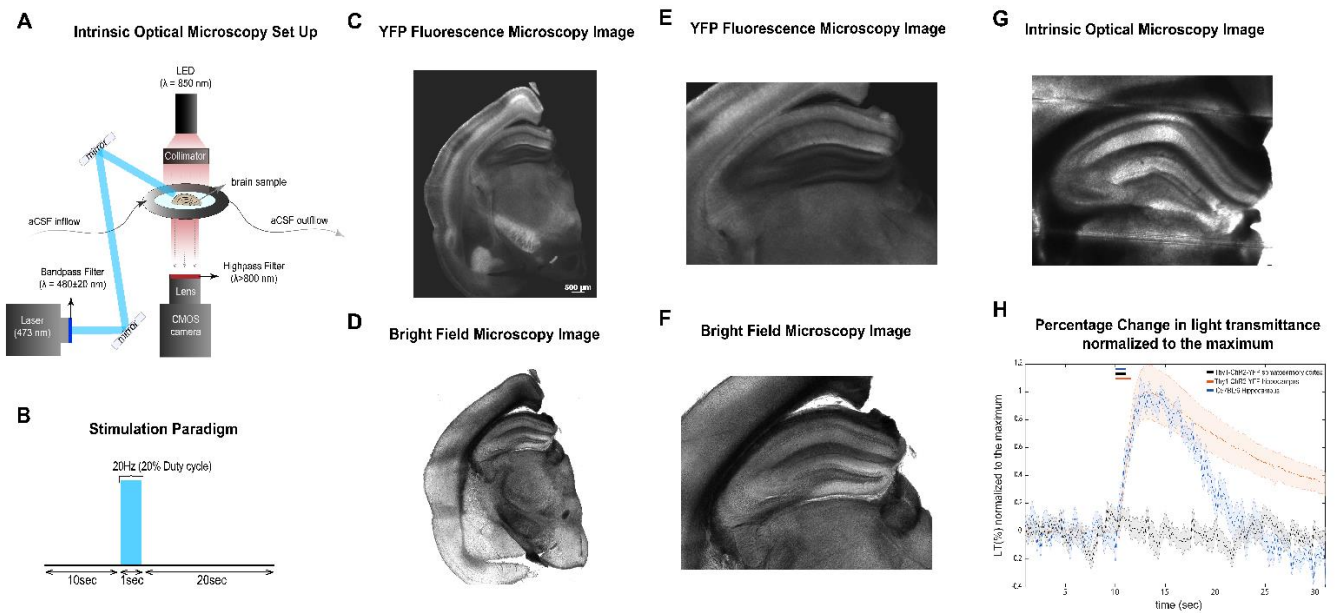

**Figure 3.** Intrinsic Optical Microscopy of acute brain slices. **(A)** Schematics of the custom-built microscope for IOM imaging, used to record activation of the CA1 region of the hippocampus in acute brain slices. **(B)** Graphical representation of the stimulation paradigm for the optogenetic experiments. **(C)** YFP fluorescence from a representative slice. **(D)** Corresponding bright field images. **(E,F)** Enlargement of the relevant areas in (C,D), respectively. **(G)** Representative images from the IOM experiments. **(H)** Changes in light transmittance averaged from the entire hippocampus due to optogenetic stimulation for hippocampal slices expressing ChR2 and control slices not expressing ChR2. Note the fast onset in these signals, reflecting the neuromorphological coupling in the ChR2-expressing slices and the absence of signals in controls; similar signals were observed in cortical areas in slices expressing ChR2.

To better compare the MRI signals with IOSs time series, and given that the signal to noise ratio of single IOM pixels is very low, we analyzed the IOM data analogously to the line scanning fMRI experiments, i.e., IOM images were averaged into a "line", to observe the IOS temporal dynamics in hippocampal layers. Figure 4A shows a schematic of the hippocampal slice with its layered structure, while Figure 4B shows a corresponding image from the functional IOM experiment. The structure of the hippocampal CA1 region (in the shadowed box of Fig. 4B) is quite uniform in the horizontal axis, thereby allowing a robust averaging across the pixels in the horizontal dimension without loss of anatomical definition (Figure 4C). In addition, CA1 exhibits a high and relatively uniform level of ChR2 expression (Figures 3C and 3D).

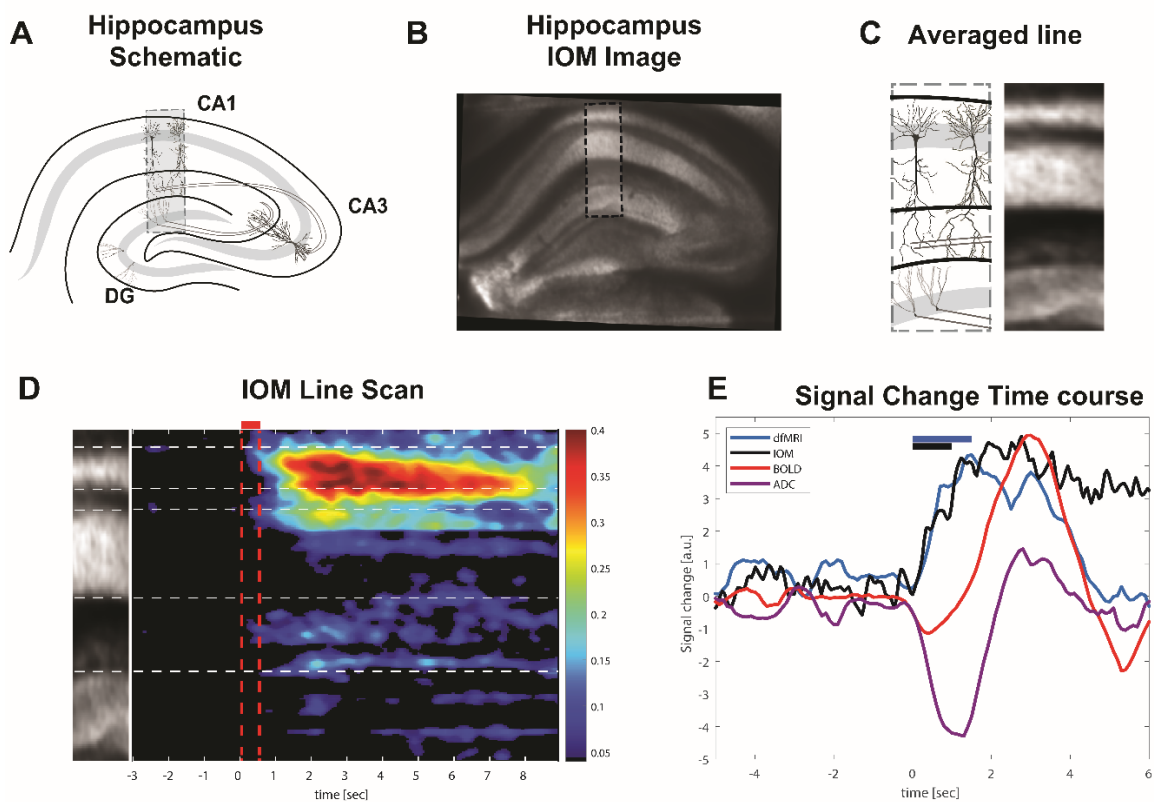

**Figure 4.** Intrinsic Optical Microscopy line scanning. **(A)** Scheme of the hippocampal slice. **(B)** image from the IOM experiments. **(C)** the pixels averaged for the IOM line scanning, along the vertical direction. **(D)** The temporal dynamics of the IOM line scanning, revealing rapid onset functional signals in areas rich with cellular processes. **(E)** A comparison of the dynamics in the MRI and IOM experiments. Note the similar rise times for dfMRI and IOSs.

Figure 4D shows the time courses obtained for a representative Thy1+ slice in such a line scanned functional IOM experiment. Importantly, rapid-onset signals could be observed in hippocampal layers followed by a slower signal that peaks between 2-3 seconds and remains high for ~20 seconds (c.f. Figure 3G). The strongest signals are observed in areas rich with axons and dendrites and on the borders where they meet cell bodies.

Finally, we compared between the time courses of dfMRI, BOLD, and ADC signals (averaged from cortical column layers II to V) and the optogenetically evoked functional IOM experiment signals (averaged from the higher levels of the hippocampus where signal was strongest). The MRI-driven dynamics mirror the results in Figure 2 and show how BOLD signals are delayed compared to the rapid-onset dfMRI (and ADC) signals. Strikingly, however, we find that the early dfMRI signals tracked the early functional IOM signals for at least ~500 ms, and perhaps even 1000 ms if some leniency is allowed. The dfMRI signals then deviate from the timecourse of the IOSs. As expected, the positive BOLD responses clearly did not track these IOS dynamics.

## Discussion

Neurovascular couplings have been at the heart of fMRI since its inception[1] and have transformed our understanding of brain function[2,49]. Still, given the limited specificity of neurovascular couplings[20], alternative coupling mechanisms have been put forth as perhaps more direct markers of neural activity. The dfMRI approach[30,32] had been previously proposed for such purposes exactly; however, its underlying mechanisms, and its coupling to neural activity have been vigorously debated[38,39,41,42]. Therefore, this study aimed to shed light into the origins of dfMRI signals, especially vis-à-vis the putative neuromorphological coupling.

Given that prior studies using IOM have shown that fast IOS components reflect cell swelling[50] associated with neuronal action potential firing, representing a "neuromorphlogical coupling", and that these couplings have a very rapid onset, we hypothesized that an ultrafast approach for dfMRI is required to detect such signals. This required the development of a new approach for dfMRI, involving the combination of large-tip-angle spin-echo sequences[48] with a line scanning[22] strategy (LTA-LS), thereby enabling high temporal resolution of diffusion-weighted signals. Perhaps the most interesting finding in this study is the discovery of the hitherto unreported rapid-onset dfMRI component evidencing increased functional signals already 100 ms post stimulation (Figure 2) and lasting several hundred of milliseconds (followed by a second component, *vide infra*). This rapid-onset dfMRI component was characterized by several noteworthy features: (1) activation was observed very rapidly after stimulus presentation, already in the first time point (100 ms) in most of the FL S1 cortex; (2) it appeared (qualitatively) quite invariant to a hypercapnia challenge (Figure S1); (3) its dynamics agreed well with the early responses of IOSs elicited optogenetically in acute brain slices devoid of BOLD components (Figure 4); finally, (4) when the dfMRI signals and SE-BOLD signals were normalized to produce ADC maps, a highly punctate response was observed, with decreases of ADC decreases temporally confined to the window of stimulation with less than 300 ms error (Figure 2E-F).

All this evidence mutually reinforces and suggests that dfMRI's rapid onset component reflects, at least to some extent, the *neuromorphological coupling* – an interaction between neural activity and (sub)cellular-scale morphological modulations on the microscopic and/or mesoscopic scales interrogatable via water diffusion. In other words, at the diffusion-weighting used, the rapid-onset dfMRI component appears to be sensitive to microstructural modulations coupled to the neural activity elicited by the stimulation. This does not mean that neurovascular couplings are completely absent from these signals, but it suggests a predominance of neuromorphological couplings in these regimes.

Many different sources could contribute to neuromorphological couplings with potentially diverse time constants and relative amplitudes. Evidence from studies using polarized-light IOM show that action potential discharge induces variations in neurite volume and bouton size due to cytoskeleton reorganization[51]. On the other hand, transmitted light IOM studies have shown that cell swelling[52], due to water influx through the cell membrane[44], and shrinkage of the extracellular space[45] strongly contribute to IOSs[44,46]. Though the main component of IOSs is of neuronal origin, astrocyte glutamate-uptake[52] and glial control of the potassium extracellular balance[53] were also implicated in these signals, with rather slow decay times. Additionally, axon diameter variations have been reported using atomic force microscopy, where a fast mechanical spike associated with action potential firing has been observed[54]. The amplitude of these mechanical spikes is around 0.5-0.8 nm[55], which could be expected to produce very small signal changes in dfMRI measurements[56]. Other mechanisms have also been reported to contribute to "cell swelling", such as transient changes in spine size[57]. Therefore, the neuromorphological coupling is here used as an "umbrella term" to collectively refer to any of these cellular mechanisms that are observed upon action potential firing. The relative sensitivity of dfMRI to each of these processes remains to be deciphered and it is still too early to associate the rapid-onset responses to a particular mechanism.

It is worth contemplating how dfMRI's early onset signals reflect neural activity. At a physiological level, vascular responses take hundreds of milliseconds to onset[5,16], although changes in deoxyhemoglobin levels can occur with ~200 ms time scale[16]. Although the slow vascular responses have been reported to be layer specific in the cortex[22], neural signals are known to activate the cortical column within a few tens of milliseconds, with inputs to layer V delayed by a few tens of milliseconds[58]. This is quite consistent with the temporal dynamics shown in Figure 2. Nevertheless, more experiments with electrophysiological or other multimodal readouts[59] combined with pharmacological manipulations are required to further investigate the neuromorphological coupling itself. The use of diffusion sequences that reduce the contribution of internal gradients[41] into the diffusion signal (consequently avoiding capturing BOLD signals in dfMRI measurements), might help retrieving purely neuromorphological couplings in dfMRI studies.

It is also interesting to try and interpret the origin of the rapid-onset dfMRI signals and the ensuing ADCs from a more biophysical perspective. The general signal attenuation in a diffusion-weighted spin-echoed sequence – such as the one used here – follows the general form of[60]

Eq. 1

$$S(t) \propto S_0 \exp(-(AG_D^2 + BG_0(t)G_D + CG_0^2(t)) * ADC(t))$$

where $S(t)$ is the temporal signal (the time index reflecting the time over the fMRI experiment), $S_0$ is a time-invariant signal scaling factor, $G_D$ is the applied diffusion gradient ($G_D = 0$ for SE-BOLD), $G_0(t)$ is the internal gradient time dependence, ADC(t) reflects the temporal dependence of the apparent diffusion coefficient due to true microstructural variations, and, *A*, *B* and *C* are sequence-dependent constants. For SE-BOLD, only the latter term contributes, while for dfMRI, all terms contribute. To calculate the ADC as in Figure 2, the log signals are normalized, i.e.,

Eq. 2

$$-\frac{\ln\left(\frac{S_{dfMRI}(t)}{S_{SE-BOLD}(t)}\right)}{AG_D^2} = ADC(t) * (1 + C'\frac{G_0(t)}{G_D})$$

where the constant C' collects (constant) nuisance terms. Note that *two* time-dependent quantities appear in Eq. 2: the microstructural ADC(t) and the purely BOLD-driven $G_0(t)$, scaled by the application of the diffusion gradient. If $G_0(t)$ is small or if $G_D$ is large ($G_D^2 \gg G_D$), the latter term will be negligible and the measured ADC(t) will mainly reflect the microstructural changes. A very useful hint toward the mechanism underlying the early dfMRI component can be viewed in the spatially averaged temporal dynamics (Figure 2B), where a small but obvious negative BOLD response appears rapidly upon stimulation (the initial dip). Despite that the initial dip is negative (i.e., the sign of $G_0(t)$ is negative), the dfMRI early onset response (Figure 2D) is positive, suggesting that ADC(t) is larger than the BOLD contribution; furthermore, the initial dip reverses its course around 300-400 ms, while the dfMRI signals continue to increase monotonically and similarly, the ADC(t) values (Figure 2F) remain monotonic. These attributes indicate that for the b-values used in this study, $G_0(t)/G_D$ is sufficiently small during the first few hundred milliseconds to render the contribution on the right hand side of Eq. 2 small and thereby suggest that the early onset dfMRI component and ADC reflects the microstructural modulations in the tissue. We also note that in a previous study, areas showing no BOLD responses still evidenced significant dfMRI signals[61], which further suggests that when $G_0(t)/G_D$ is small, the dfMRI signals can reflect neuromorphological couplings.

The rapid onset dfMRI component was clearly followed by more complex dynamics (Figure 2) as can be observed e.g., after 1000-1500 ms. This later dfMRI response has likely been the key feature detected in previous conventional dfMRI studies with temporal resolution of several seconds. Given that it appears to strongly modulate with hypercapnia (Figure S2) and given that its dynamics do not follow IOSs, it seems likely that BOLD effects are nonnegligible in this component. It should be stressed that this does not mean that microstructural effects are not present in these later dynamics; however, they may be convolved with BOLD effects (Eq. 2), as also suggested by others[38]. This probably explains why a recent study failed to detect changes in diffusivity with chemical stimulation in slices[42] – the temporal resolution was likely not sufficiently fast to detect the rapid onset component. In the future, it may be interesting to try to tease apart the relative contributions of each of these responses by performing experiments with varying $G_D$, whose linear and squared dependencies could perhaps be disentangled, thereby revealing each of the components in Eq. 2 separately.

As any other study, we recognize several limitations for our work. Perhaps foremost, is that the in-vivo aspects of this study were performed in sedated rats stimulated via sensory inputs lasting 1.5 seconds, while the IOM signals were recorded in acute physiological slices originating from the mouse and stimulated optogenetically for 1 sec. The differences in species and in stimulation modality, must be considered when interpreting the findings of this study. In addition, LTA-LS requires the use of saturation slices that may accentuate blood flow effects, especially in the SE-BOLD signals. Still, our findings are encouraging for future studies potentially overcoming some of these limitations.

In conclusion, using an ultrafast scanning approach, we have shown that dfMRI signals carry a rapid-onset component (~100 ms) that appears to be rather insensitive to a vascular challenge and whose dynamics track those of IOSs in acute brain slices, which are thought to represent microstructural and mesoscopic modulations associated with action potential firing. Therefore, we conclude that this component reflects a neuromorphological coupling, whose noninvasive detection could potentially be important in mapping neural activity with mechanisms different from the neurovascular couplings. These features bode well for future applications of dfMRI in health and disease.

# Methods

All animal experiments followed the experimental procedures in agreement with Directive 2010/63 of the European Parliament and of the Council, and all the experiments in this study were preapproved by the Champalimaud Animal Welfare Body and the national competent authority (Direcção Geral de Alimentação e Veterinária, DGAV), under the approved protocol 0421/000/000/2016.

**Animal preparation.** Long Evans male rats (n = 16) 8-10 weeks old were housed in pairs, with a light/dark cycle of 12 h. All in-vivo experiments were performed under sedation. Briefly, rats were induced into deep anesthesia in a custom cage with 5% isoflurane (Vetflurane, Virbac, France). Once sedated, the rats were placed in a custom MRI animal bed (Bruker Biospin, Karlsruhe, Germany) and maintained under ~2.5% isoflurane while being fixed using a bite bar and ear bars. At this stage, two stimulation electrodes were inserted into the left forepaw between digits 1-2 and 4-5, and the animals were switched to a subcutaneous medetomidine sedation protocol[62] (Dormilan, Vetpharma Animal Health, Spain) consisting of 1 mg/ml solution diluted 1:10 in saline. A 0.05 mg/kg bolus was injected and upon 5 min constant infusion of 0.1 mg/kg/h delivered via a perfusion pump (GenieTouch, Lucca Technologies, USA) was started, while stepwise decreasing the isoflurane concentration in breathing air (Medical Air, Linde Healthcare Portugal). This preparation procedure usually took ~10 min; functional MRI experiments were not commenced before ~30 min had passed after the isoflurane was removed from the breathing air. A rectal temperature probe and respiration sensor were also used to continuously monitor the temperature and respiration rate (Model 1025, SAM-PC monitor, SA Instruments Inc., USA), respectively, of rats during the experiment. In hypercapnia experiments, $pCO_2$ was monitored using a transcutaneous monitoring system (TCM4 series, Radiometer, Denmark). Sedation was reverted at the end of each experiment, by injecting subcutaneously the same amount of the initial bolus of atipamezole 5 mg/ml solution diluted 1:10 in saline (Antisedan, Vetpharma Animal Health, Spain).

**In vivo MRI experiments.** All in-vivo MRI experiments were performed on a 9.4 T Bruker BioSpec scanner operating an AVANCE III HD console and using a gradient system capable of producing up to 660 mT/m isotropically. An 86 mm volume was used for transmission and ensured a relatively uniform $B_1$ profile, while a 4-element array receive-only cryogenic coil (Bruker BioSpin, Fallanden, Switzerland) was used for signal reception[63].

**Anatomical references:** During the change of anesthesia from isoflurane to medetomidine, anatomical reference images were acquired and necessary adjustments were performed. Briefly, a $B_0$ map was acquired using ParaVision 6.01's routines, and the field map was calculated. Anatomical references were acquired using a FLASH sequence (FLASH; TR/TE = 60/2 ms, flip angle=15 degrees, field of view (FOV) = 25.6x25.6 mm$^2$, spatial resolution = 0.2x0.2mm$^2$, slice thickness =1 mm) for 3D positioning of the animal head and coronal views of the brain were acquired using a T2 RARE sequence (TR/TE/TE$_{eff}$ = 1800/8/32 ms, RARE factor = 8, FOV = 16x16 mm$^2$, spatial resolution = 0.1x0.1 mm$^2$, slice thickness = 0.75 mm) to subsequently position the line for functional acquisitions.

**Stimulation paradigm:** A stimulator built in-house (Dexter Electrostimulator 1.0, Hardware Platform, Champalimaud Foundation) was used to generate square waveforms for electrical stimulation (Figure 1A). The stimulation protocol consisted of 40 seconds of rest, followed by 1.5 seconds stimulation with electrical pulses delivered to the left forepaw with a square waveform comprising 1.5 mA, 10 Hz and 3 ms stimulus duration. A total of 80 stimulation periods per experiment were used (Figure 1A).

**BOLD fMRI and dfMRI line-scanning experiments:** All functional experiments were preformed using the same LTA-LS pulse sequence, with identical acquisition parameters. For dfMRI experiments, a pair of diffusion sensitizing gradients imparted a diffusion weighting of b = 1.5 ms/μm$^2$ (Δ/δ = 16/2.2 ms). The BOLD fMRI experiments simply set the b-value to 0 ms/μm$^2$, leading to the equivalent SE-BOLD fMRI acquisition. Common pulse sequence parameters were as follows: TR/TE = 100/24 ms, FOV = 5.8 mm (1D, no phase encoding), matrix size = 68, line resolution = 85 µm, slice thickness = 1 mm, saturation bands FOV= 10 mm.

**Pulse sequence:** Diffusion MRI experiments were performed using spin echo sequences, to refocus T2*-related decay. However, due to the refocusing pulse(s), such sequences are not directly compatible with fast acquisitions due to strong T1 weighting. Here, we harnessed Large Tip Angle (LTA) approaches[48], and simply replace the 90 degree excitation pulse with a 150 degree pulse (this angle was empirically selected to maximize signal to noise ratio at a repetition time (TR) of 100 ms). Line scanning was achieved by removing the phase encoding dimension, and rather using saturation bands to select the line of interest as first proposed in Yu et al. for gradient-echo pulse sequences[22].

**Functional analysis.**

All functional MRI data, regardless of whether BOLD or dfMRI, underwent the same analysis pipeline. Data were reconstructed taking into account the sensitivity profile of the four reception coils. Outliers were automatically detected and removed (<0.16% of data were identified as outliers, these few points were then interpolated). The data was then denoised using Marchenko-Pastur Principal Component Analysis denoising[64]. A bandpass filter was then applied to the data, with 0.02 and 4.75 Hz stopband frequencies and 0.04 and 4.25 Hz passband frequencies (equiripple design method). The 2D [spatial temporal] data were converted to percent change with respect to the median signal in the rest period. A maximum likelihood estimation over trials and then over animals ensued for every pixel and timepoint. ADCs were calculated from the raw data using $ADC(t) = -\frac{1}{b}\ln(\frac{S_{dfMRI}}{S_{BOLD}})$ followed by all steps mentioned above, except that a [4 4] median filtering was added before analysis to enable a more robust estimation of ADC. Finally, for presentation purposes, data were detrended and spatiotemporally smoothed using a gaussian filter with σ=2. Where indicated, temporal profiles were extracted from the first five layers (with highest SNR) by summing the individual contribution in every pixel.

**Slice preparation:** Acute hippocampal coronal slices (300 μm thick) were prepared from 6-8 weeks old transgenic mice Thy1-ChR2-YFP$^{+/-}$ (stock no. 007615, The Jackson Laboratory, USA) or control C57Bl6/j mice (Charles River, France), using a vibratome (Leica VT1200S; Leica Biosystems, Germany) while submerged in ice-cold oxygenated Ringer slicing solution (in mM): 125 NaCl, 2.5 KCl, 25 NaHCO$_3$ and 1.25 NaH$_2$PO$_4$, 3 myo-Inositol, 2 Na-pyruvate, 0.4 ascorbic acid, 0.1 CaCl$_2$, 3 of MgCl$_2$. Slices were transferred and incubated ~30 min in a 37 ºC warm oxygenated bath solution (in mM): 125 NaCl, 2.5 KCl, 25 NaHCO$_3$ and 1.25 NaH$_2$PO$_4$, 1 MgCl$_2$, 2 CaCl$_2$. Osmolarity of both solutions was measured with a vapor pressure osmometer (Wescor Inc., USA) and adjusted to 310 ± 5 mOsm using glucose ( typically 20-25 mM). All chemicals were ordered from Sigma-Aldrich, USA. During the experiment, the slice incubation chamber was filled and constantly perfused with 37 ºC warm oxygenated bath solution.

**Intrinsic Optical Microscopy (IOM) set up:** The IOM set up was custom built in-house (Figure 3A). An inverted microscope was built where samples were illuminated form the top with an infrared LED (λ=850 nm; Thorlabs Inc., USA) and transmitted light was recorded from the bottom with a PointGrey Grasshoper3 camera (GS3-U3-32S4M FLIR Systems Inc., USA) equipped with a NIKON AF-NIKKOR 35mm f/2D lens (NIKON, Japan). A light collimator (Thorlabs Inc., USA) was placed between the sample and the LED in order to parallelize the light beams before reaching the sample. Tissue samples were placed in a 37°C heated bath chamber (TC-344C, Warner Instruments, USA) with constant flow of bath solution at a rate of 8 ml/min. A blue laser (λ=473nm, Thorlabs Inc., USA) was used for optogenetic sample stimulation and, two tilted mirrors (PF10-03-P01, Thorlabs Inc., USA) were used to direct the laser beam onto the sample. Since the laser beam arrived to the sample from the side, the amount of laser light directed to the camera was minimized. In order to further confine the light reaching the lens, two filters were added to the system: a λ = 480±20 nm bandpass filter (Thorlabs Inc.,USA) was placed at the exit of the laser in order to restrict the wavelengths of stimulation and a λ > 800 nm high pass filter was placed before the lens to eliminate any lower wavelengths sources from being captured by the camera. In addition, both the lens and the camera were covered with aluminum foil to further isolate this part of the system from ambient light. The entire setup was placed in a dark chamber during the experiments.

**IOM acquisitions:** Image acquisition was performed using the FlyCapture software (FLIR Systems Inc., USA) with FOV = 976 x 670 pixels (approximate resolution of 3x5 µm$^2$ per pixel) acquired at 20.3 Hz. A BONSAI routine[65] was developed to record the data and online data visualization. A custom-built software using the open-source Arduino software (IDE; https://www.arduino.cc) was used to control an Arduino MEGA2560 that initiated the camera and the data acquisition routine as well as it controlled the laser (Model MBL-FN-473-50, Ultralasers Inc., USA) during stimulation (frequency of stimulation=20 Hz, pulse width=10 msec, stimulation time=1 sec, laser intensity at fiber tip = 5 mW ). Each trial had a total duration of 90 sec but only 20 sec after stimulation were used for display proposes (Figure 3B). For the cortical slices, N = 9 acute cortical slices were obtained and underwent the same procedure as above for the hippocampal slices, except for a stimulation duration of 1.5 sec every 45 seconds; one slice was excluded from analysis due to artifacts. At the end of each experiment the brain slices imaged were stored in 4% PFA to control for YFP expression. Bright field images and epifluorescence images were acquired using a Zeiss AxioImager M2 microscope (Zeiss, Germany) with a 10x plan-apochromat lens (NA=0.45; Zeiss, Germany) or a 20x plan-apochromat lens (NA=0.8; Zeiss, Germany), equipped with a digital camera (Orca-Flash4.0LT, Hamamatsu Photonics, Japan).

**Intrinsic Optical Microscopy data analysis:** Raw data was analyzed using a home-written code in Matlab (The MathWorks Inc., USA). In each imaged hippocampus a region of interest (ROI) was manually drawn and the signal from all voxels was averaged. Following detrending of the signal, ROI and the percent change was computed. Two slices were excluded from the analysis because no robust activation was found, likely due to being recorded ~4 hrs after the slicing procedure. For the line scanning IOM approach, every image in the time series was equally rotated using Matlab's *imrotate* function with bilinear interpolation to ensure that CA1 was placed horizontally. The CA1 area was cropped and its short dimension was collapsed into 1D by averaging all the pixels in that dimension. Each pixel in the ensuing CA1 "line" was then subject to the same analysis as the ROI data, namely, empirical mode decomposition detrending and conversion to percentage change. For comparison of MRI and IOM data, all time-courses were z-scored and aligned along the vertical axis to enable a fair comparison.

## Acknowledgements

This study was funded in part by the European Research Council (ERC) (agreement No. 679058), as well as by Fundação para a Ciência e Tecnologia (Portugal), project 275-FCT-PTDC/BBB-IMG/5132/2014. The authors acknowledge the vivarium of the Champalimaud Centre for the Unknown, a facility of CONGENTO which is a research infrastructure co-financed by Lisboa Regional Operational Programme (Lisboa 2020), under the PORTUGAL 2020 Partnership Agreement through the European Regional Development Fund (ERDF) and Fundação para a Ciência e Tecnologia (Portugal), project LISBOA-01-0145-FEDER-022170. The authors would like to thank Dr. Jelle Veraart for assistance in data reconstruction and helpful discussions.

# Figures and Tables

**Figure 1.** Large tip-angle line scanning data acquisition. (A) Schematic representation of the stimulation paradigm in sedated rats. Stimulation needles in the left forepaw delivered electrical square pulses to induce temporally precise activation of the forelimb primary somatosensory cortex (FL S1). (B) Schematic representation of the pulse sequence used to acquire LTA-LS data. A pulsed gradient spin-echo sequence compromised a saturation band module, a tip-angle $\theta_{LTA}$=155 degrees, followed by a $\pi$ pulse. A slice selective gradient ($G_{slice}$) was used to spatial encoding and readout gradients ($G_{RO}$) for frequency encoding during acquisition (ACQ). Diffusion sensitizing gradients ($G_{diffusion}$) were applied on each side of the $\pi$ pulse to impart diffusion weighting in dfMRI measurements and were turned-off during SE-BOLD measurements. (C) Diagram of the line scanning method, showing the positioning of the saturation slices.

…

Abbreviations: DRG – dorsal root ganglion; DCN – dorsal central nucleus (dorsal column-medial lemniscus pathway); FL S1 – forelimb primary somatosensory cortex.

**Figure 2.** Temporal profiles of FL S1 activation. (A) SE-BOLD, (B) dfMRI and (C) ADC temporal activation maps, in percentage, of the FL S1 region. In the left, the cortical depth denotes the FL S1 column depth and in the right cortical layers I to VI are denoted, highlighted by white dashed horizontal lines. The red vertical solid bars denote stimulation time. (D) SE-BOLD, (E) dfMRI and (F) ADC spatially averaged time courses of the data, in percentage change. The red horizontal bars denote stimulation.

**Figure 3.** Intrinsic Optical Microscopy of acute brain slices. (A) Schematics of the custom-built microscope for IOM imaging, used to record activation of the CA1 region of the hippocampus in acute brain slices.

(B) Graphical representation of the stimulation paradigm for the optogenetic experiments. (C) YFP fluorescence from a representative slice. (D) Corresponding bright field images. (E,F) Enlargement of the relevant areas in (C,D), respectively. (G) Representative images from the IOM experiments. (H) Changes in light transmittance averaged from the entire hippocampus due to optogenetic stimulation for hippocampal slices expressing ChR2 and control slices not expressing ChR2. Note the fast onset in these signals, reflecting the neuromorphological coupling in the ChR2-expressing slices and the absence of signals in controls; similar signals were observed in cortical areas in slices expressing ChR2.

**Figure 4.** Intrinsic Optical Microscopy line scanning. (A) Scheme of the hippocampal slice. (B) image from the IOM experiments. (C) the pixels averaged for the IOM line scanning, along the vertical direction. (D) The temporal dynamics of the IOM line scanning, revealing rapid onset functional signals in areas rich with cellular processes. (E) A comparison of the dynamics in the MRI and IOM experiments. Note the similar rise times for dfMRI and IOSs.